\documentstyle[twoside,fleqn,espcrc2,epsf]{article}

\newcommand{\AmS}{{\protect\the\textfont2
  A\kern-.1667em\lower.5ex\hbox{M}\kern-.125emS}}

\title{Lattice QCD on a Beowulf Cluster}

\author{Seyong Kim\address{Department of Physics, Sejong University,
        Seoul, Korea}\thanks{The author thanks Samsung Electronics
        Co. for the support for this project.}}
       
\begin{document}

\begin{abstract}

Using commodity component personal computers based on Alpha processor
and commodity network devices and a switch, we built an 8-node
parallel computer. GNU/Linux is chosen as an operating system and
message passing libraries such as PVM, LAM, and MPICH have been tested
as a parallel programming environment.  We discuss our lattice QCD
project for a heavy quark system on this computer.

\end{abstract}

\maketitle


\begin{figure}
\begin{center}
\epsfxsize=60mm
\leavevmode
\epsffile[161 250 452 552]{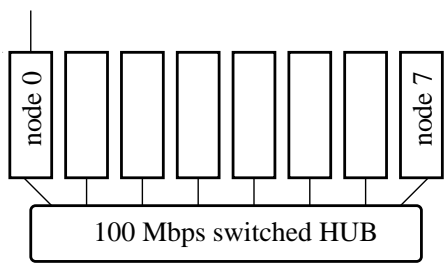}
\caption{Network configuration of the cluster}
\label{fig:netconfig}
\end{center}
\end{figure}

Even a modest lattice QCD project demands quite large amount of
computing resources. In this regard, it has always been an attractive
idea to build a cheap high performance computing platform out of
commodity PC's and commodity networking devices. However, the
availability of cheap hardware components solved only a part of
problem in building a parallel computer in the past. There were large
hidden cost in constructing a do-it-yourself parallel computer and
only groups which could dedicate significant amount of resources were
able to take advantage of this idea. Chief stumbling block has been in
providing parallel programming environment (both in hardware and
software) from the scratch and in maintaining one of a kind
hardware. Following recent trend in do-it-yourself clustering
technology \cite{Beowulf}, we built a cluster which uses only
available hardware and software components and can be easily
maintainable. Here, we discuss our experience.


In terms of hardware, the node level configuration of our cluster does
not differ from ordinary PC's other than the fact that it is
monitor-less. Each node consists of a single 600 MHz Alpha 21164
processor and SDRAM SIMM main memory. The amount of memory on
individual nodes varies from 128 Mbytes (5 nodes) to 256 Mbytes (2
nodes). SCSI hard disks on each nodes has either 2 Gbytes (4 nodes) or
4 Gbytes (4 nodes). Additionally, each node has CD-ROM drive and 3 1/2
inch floppy drive. Power requirement of each node is 300 Watt.  As a
network component, each node has a 100 Mbps Ethernet card (3Com
3C905). Node 0 which serves as a front-end has one more
100 Mbps Ethernet card for outside connection. For the inter-processor
communication, we use a 100 Mbps switched HUB (24 port Intel
510T). Unlike the bus structure of a HUB, this inexpensive device
allows simultaneous communications among the nodes and offers a
flexibility in communication topology.

\begin{figure}
\begin{center}
\epsfxsize=60mm
\leavevmode
\epsffile[161 250 452 552]{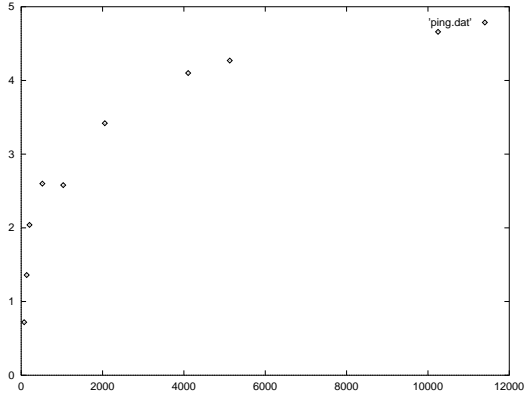}
\caption{Network bandwidth (ping test). The vertical axis is
Mbytes/sec and the horizontal axis is ping data size in bytes}
\label{fig:ping}
\end{center}
\end{figure}

\begin{figure}
\begin{center}
\epsfxsize=60mm
\leavevmode
\epsffile[161 250 452 552]{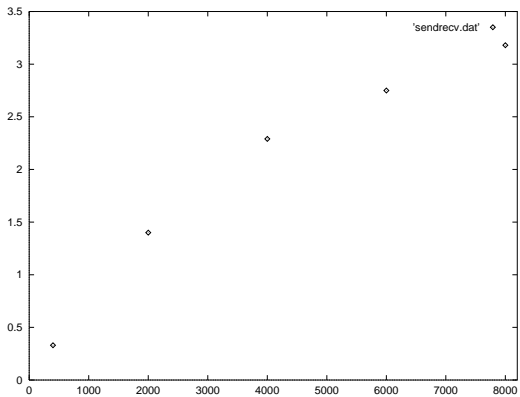}
\caption{Network bandwidth (round robin test). The vertical axis is
Mbytes/sec and the horizontal axis is message size in bytes}
\label{fig:sendrecv}
\end{center}
\end{figure}

Fig. \ref{fig:netconfig} shows the network configuration of our
cluster. Since we use a switch, the communication distance between any
two nodes are the same unless the number of nodes becomes larger than
the number of available ports in the switch. To outside world, only
node 0 exists. All the nodes are assigned local subnet addresses
(192.168.1.1 -- 192.16.1.8) where 198.168.x.x are reserved addresses
specifically for a private subnet and node 0 acts as a gateway to the
rest. In this way, we can increase the number of nodes in
the cluster without worrying about available IP addresses.
As a node operating system, we use Alzza Linux version 5.2a
\cite{Alzza} for an Alpha processor which is a Korean customized
version of Red Hat Linux 5.2 \cite{RedHat} with kernel version 2.2.1.
Three different parallel programming environments, LAM (Local Area
Multicomputer) version 6.1 \cite{LAM}, MPICH (Message Passing
Interface-Chameleon) version 1.2.2 \cite{MPICH}, and PVM (Parallel
Virtual Machine) version 3.4 \cite{PVM} have been tested on our
platform. These are all based on the message passing paradigm of
parallel computing and use TCP/IP mechanism for the actual
communication. Linux comes with FORTRAN and C compiler and the
parallel programming environments offer wrappers for these languages.
Since these parallel programming environments use remote shell (rsh)
for a parallel job execution, users need to have accounts on each
nodes. NIS system is used for the password validation. Hard disk space
on each nodes has divided into three different partitions : one for
local operating system, the other for NFS mounted '/home' directory
and the third for a scratch space for large I/O operations.
Installation procedure consists of two parts : one for Linux operating
system setup and the other for parallel programming library
setup. Once Internet setup for each node is properly done, subnet
network can be established by just connecting Ethernet ports.
The overall cost for building our 8 node configuration is shown in
Table \ref{tab:cost}. Cost for the console device such as a monitor,
mouse and keyboard are not included since we use a used one (this
table should be taken as a rough indication for the cost of our
cluster since the component price changes quite rapidly).

\begin{table*}
\setlength{\tabcolsep}{1.5pc}
\newlength{\digitwidth} \settowidth{\digitwidth}{\rm 0}
\catcode`?=\active \def?{\kern\digitwidth}
\caption{Cost for our 8 node cluster}
\label{tab:cost}
\begin{tabular}{lcrr}
\hline
        \multicolumn{1}{c}{component} &
        \multicolumn{1}{c}{no. of components} &
        \multicolumn{1}{r}{price/component} &
        \multicolumn{1}{r}{Net price} \\
\hline\hline
PC (memory, HDD and etc) & 8&$\sim 3000 \$$ &$\sim 24,000 \$$ \\
additional LAN card  & 1  &  70 \$ &  70 \$ \\
LAN cable & 9  & - & 35 \$ \\
switch    & 1  & 2100 \$ & 2100 \$ \\
UPS (3Kwatt)  & 1  & 1560 \$ & 1560 \$ \\
\hline\hline
total     &    &    & 27,660 \$\\
\hline
\end{tabular}
\end{table*}


Since performance of a cluster is determined by (single node
performance $-$ system overhead due to inter-node communication)
$\times$ the number of nodes, sustained speed of a single CPU and
efficiency of network component play an important role in a cluster.
Under GNU/Linux compiler, various tests showed that the sustained
speed of a single Alpha processor is better than that of an Intel
processor just by the difference in CPU clock speed. It is because
Alpha 21164 processor does not support out of order execution (under
the same condition, Alpha 21264 which supports out of order execution
does better than Alpha 21164 by about factor two). Serial version of
our quenched code for an $8^3 \times 32$ lattice which is coded with
$SU(3)$ index as the inner-most loop and uses multi-hit and
over-relaxation algorithm achieved 50 MFLOPS. Under Compaq FORTRAN
compiler for Linux system (beta version), the same code achieved 91
MFLOPS (a code with long inner-most loop may do better under Compaq
FORTRAN compiler by factor 4 or more \cite{Ryu}). In contrast, the
same code on a 200 MHz Intel Pentium II MMX achieved 18 MFLOPS under
GNU/Linux compiler. This single node benchmark suggests that with the
same device, we can take advantage of future development of compiler
without further tuning of codes as GNU/Linux compiler improves.  As
for the network performance, we tested the network setup using two
different methods. One is using ``ping'' test and the other is using
``round-robin'' communication. ``ping'' uses ICMP layer on top of IP
layer and ``round-robin'' test uses TCP layer on top of IP layer.
Fig. \ref{fig:ping} shows ``ping'' test bandwidth and
Fig. \ref{fig:sendrecv} shows ``round-robin'' test bandwidth of LAM
parallel programming environment. We found that LAM does better than
MPICH for short message and MPICH does better than LAM for large
message. Although we have a dedicated network for our cluster system,
three parallel programming environments we have tested all assume
normal LAN environment and use TCP/IP layer before the link layer in
order to avoid various problems from sharing communication
network. Further improvement in communication speed can be achieved if
UDP layer with error handling is used instead of TCP. Under GNU/Linux,
MPI parallel version of our quenched code for a $16^3 \times 32$
lattice achieved 346 MFLOPS with LAM and 378 MFLOPS with MPICH. Thus,
communication overhead is about 21 \% for LAM and 14 \% for
MPICH. Paralle code has not been tested under Compaq FORTRAN compiler
yet.


Currently, we are generating full QCD configurations on a $8^3 \times
32$ lattice at $\beta = 5.4$ with $m_q a = 0.01$ for heavy quark
physics and we found that relatively cheap high performance computing
platform can be easily constructed and maintained using all commodity
software and hardware.


\begin{thebibliography}{9}
\bibitem{Beowulf} Beowulf project,
 http://cesdis.gsfc.nasa.gov/linux/beowulf/\\ beowulf.html
\bibitem{Alzza} LinuxKorea Corp., http://www.korealinux.co.kr/.
\bibitem{RedHat} Red Hat Software, inc., http://www.redhat.com/.
\bibitem{LAM} LAM/MPI Parallel Computing, http://www.mpi.nd.edu/lam.
\bibitem{MPICH} MPICH - A Portable MPI implementation,
http://www.mcs.anl.gov/mpi/mpich/index.html.
\bibitem{PVM} PVM: Parallel Virtual Machine, http://www.epm.ornl.gov/pvm.
\bibitem{Ryu} D.S. Ryu, private communication.
\end{thebibliography}
\end{document}